%% file: Paper-0035.tex
\documentclass[runningheads]{llncs}
\usepackage[T1]{fontenc}
\usepackage{fontawesome}
\usepackage{svg}
\usepackage{multirow}
\usepackage{graphicx}
\usepackage[hidelinks]{hyperref}
\usepackage[numbers]{natbib}
\begin{document}
%
% \title{AxonDeepSynth: Histology Dataset Translation for Out-of-Distribution Segmentation Pseudo labeling}
% \title{Unpaired Modality Translation for Label-Free Domain Adaptation in Histological Axon and Myelin Segmentation}
\title{Unpaired Modality Translation for Pseudo Labeling of Histology Images}
%
%Histology Pseudo Labeling via Unpaired Modality Translation
\titlerunning{Histology Image Pseudo Labeling via Modality Translation}
\author{Arthur Boschet\inst{2} \and Armand Collin \inst{1,2} \and
Nishka Katoch\inst{1} \and Julien Cohen-Adad\inst{1,2}}
%index{Boschet, Arthur} 
%index{Collin, Armand} 
%index{Katoch, Nishka} 
%index{Cohen-Adad, Julien} 

\authorrunning{A. Boschet et al.}
\institute{NeuroPoly Lab, Institute of Biomedical Engineering, Polytechnique Montréal, Montréal, Québec, Canada \and
Mila - Québec Artificial Intelligence Institute, Montréal, Québec, Canada}

\maketitle              % typeset the header of the contribution
\begin{abstract}
The segmentation of histological images is critical for various biomedical applications, yet the lack of annotated data presents a significant challenge. We propose a microscopy pseudo labeling pipeline utilizing unsupervised image translation to address this issue. Our method generates pseudo labels by translating between labeled and unlabeled domains without requiring prior annotation in the target domain. We evaluate two pseudo labeling strategies across three image domains increasingly dissimilar from the labeled data, demonstrating their effectiveness. Notably, our method achieves a mean Dice score of \(0.736 \pm 0.005\) on a SEM dataset using the tutoring path, which involves training a segmentation model on synthetic data created by translating the labeled dataset (TEM) to the target modality (SEM). This approach aims to accelerate the annotation process by providing high-quality pseudo labels as a starting point for manual refinement.
% ANONYM
% \footnote{https://axondeepseg.readthedocs.io/}.

\end{abstract}
\section{Introduction}
Histology images play a crucial role in the study of neurodegenerative disorders, such as Parkinson's \cite{Signaevsky2022} or Alzheimer's disease \cite{McKenzie2022}. High resolution histology imaging of neurological tissues can be obtained using scanning (SEM) or transmission (TEM) electron microscopy, bright field imaging (BF) or Coherent anti-Stokes Raman spectroscopy (CARS), each producing images of different resolution and contrasts, making it difficult to train a segmentation model that works across a variety of imaging technologies and acquisition parameters. Recent advances in deep learning have largely improved models to segment microscopy images \cite{zaimi_axondeepseg_2018, litjens_survey_2017}. For instance, the U-Net architecture \cite{ronneberger_u-net_2015}, a CNN-based model has become a standard in biomedical image segmentation due to its ability to produce precise segmentation with the help of annotated data \cite{isensee_nnu-net_2021}. Unfortunately, one challenge in developing effective models of biomedical segmentation is the low amount of labels, which are modality specific. Annotating histology images is a labor and time-intensive process that requires expert knowledge. A new annotation strategy is to use image translation techniques to augment existing datasets and generate synthetic annotated images. With translations often unpaired, traditionally generative adversarial networks (GANs) \cite{goodfellow2014generative} have been used. Diffusion models \cite{ho2020denoising} have recently emerged as a robust alternative, based on explicit likelihood characterization and a gradual sampling process. We utilize SynDiff \cite{ozbey2023unsupervised}, an adversarial diffusion model that uses conditional diffusion for efficient modality translation.

Our main contribution is a microscopy pseudo labeling pipeline based on unsupervised image translation. The proposed weakly supervised method generates pseudo labels by image translation between the labeled and unlabeled data, and does not require any prior annotation in the target domain. This contrasts with transfer learning, which is the most prevalent approach to mitigate data scarcity in medical deep learning. Unlike \cite{Xing2019}, where image translation was used to perform pseudo labeling for object detection in microscopy images, we explore this idea for a semantic segmentation task and compare two pseudo labeling strategies on three increasingly challenging target datasets, while providing recommendations as to when to deploy this pipeline for optimal results. The first strategy trains a proxy segmentation model on synthetic data derived from modality-translated labeled images, while the second translates unlabeled images to match a labeled data distribution, enabling segmentation with a pre-trained model. Our code is available online \footnote{https://github.com/axondeepseg/AxonDeepSynth}.

% ANONYM
%\footnote{https://github.com/******}. The model is also directly integrated into the AxonDeepSeg software, for a user-friendly experience with access to morphometrics extraction tools.

\input{methods}

\input{results}

\section{Conclusion}
Our results provide a better understanding of the optimal scenarios where leveraging image translation is advantageous for pseudo-labeling. When the target unlabeled dataset is sufficiently similar to the available labeled data, our pseudo labeling strategies do not provide a significant advantage over applying a pre-trained model, as demonstrated in the TEM $\leftrightarrow$ TEM-MACAQUE experiment. However, in both the TEM $\leftrightarrow$ SEM and TEM $\leftrightarrow$ BF scenarios, the target dataset is dissimilar enough that applying a pre-trained model is ineffective. In these cases, our pseudo labeling strategies are effective and provide useful initial masks to be manually corrected instead of annotated from scratch, saving precious time in the process. Realistically, a Dice score superior to 0.5 goes a long way, reducing the annotation time by 25\%-50\% (although this should be further quantified in future works). We recommend first using the adaptive path which does not require training an intermediate segmentation model. The tutoring path can provide a complementary set of pseudo labels, at the additional cost of training a proxy segmentation model. We hope this works inspires other groups to recycle available datasets when applicable to acquire more labeled data.

\begin{credits}
\subsubsection{\ackname} We would like to thank Simeon Christian Daeschler, Marie-Hélène Bourget, Tessa Gordon, Gregory Howard Borschel, Charles R. Reiter, Geetanjanli Bendale and Osvaldo Delbono for the BF dataset.
% ANONYM
% \subsubsection{\ackname} We would like to thank Tanguy Duval and Daniel Côté for the CARS images, Simeon Christian Daeschler, Marie-Hélène Bourget, Tessa Gordon and Gregory Howard Borschel for the \texttt{BF1} dataset, Charles R. Reiter and Geetanjanli Bendale for the \texttt{BF2} dataset, and Osvaldo Delbono for the \texttt{BF3} dataset. 
% (SOMETHING ELSE TO ADD?)

\subsubsection{\discintname}
The authors have no competing interests to declare that are relevant to the content of this article.
% It is now necessary to declare any competing interests or to specifically
% state that the authors have no competing interests. Please place the
% statement with a bold run-in heading in small font size beneath the
% (optional) acknowledgments\footnote{If EquinOCS, our proceedings submission
% system, is used, then the disclaimer can be provided directly in the system.},
% for example: The authors have no competing interests to declare that are
% relevant to the content of this article. Or: Author A has received research
% grants from Company W. Author B has received a speaker honorarium from
% Company X and owns stock in Company Y. Author C is a member of committee Z.
\end{credits}

\bibliographystyle{splncs04}
\bibliography{Paper-0035}

\end{document}

\section*{Appendix}

\renewcommand{\thefigure}{S1}
\begin{figure}[H]
\centering
\includegraphics[width=0.8\textwidth]{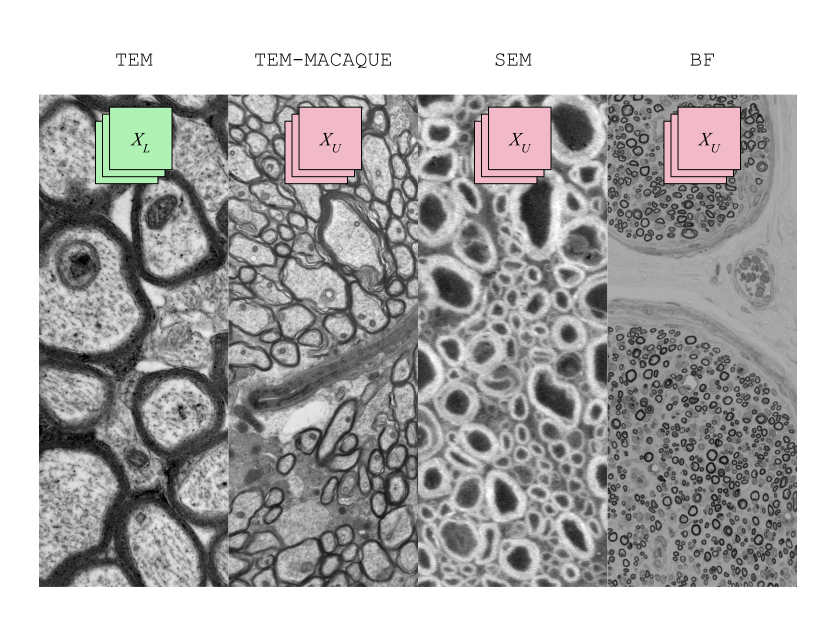}
% \includesvg[width=1.0\textwidth]{figs/figure_dataset_overview.svg}
\caption{Datasets used, either annotated ($X_L$) or unannotated ($X_U$). The \texttt{TEM} dataset is recycled for pseudo labeling in all three unlabeled domains.} \label{fig2-data}
\end{figure}

\subsubsection{Modality Translation Training}

The SynDiff framework translates images between distributions \(A\) and \(B\). It includes four modules: two non-diffusive modules \((G_{\phi^A}, D_{\phi^A})\) and \((G_{\phi^B}, D_{\phi^B})\), and two diffusive modules \((G_{\theta^A}, D_{\theta^A})\) and \((G_{\theta^B}, D_{\theta^B})\). Images \(x_0^A \in A\) and \(x_0^B \in B\) are transformed as follows:

\begin{equation*}
\scalebox{0.8}{$
    \tilde{y}^A = G_{\phi^A}(x_0^B), \quad \tilde{y}^B = G_{\phi^B}(x_0^A), \quad \tilde{x}^A_0 = G_{\theta^A}(x^A_t, y=\tilde{y}^B, t), \quad \tilde{x}^B_0 = G_{\theta^B}(x^B_t, y=\tilde{y}^A, t)
$}
\end{equation*}

\paragraph{Adversarial Loss for Non-Diffusive Modules:} The non-diffusive modules use non-saturating adversarial losses:
\begin{equation*}
\scalebox{0.8}{$
\begin{split}
    L_{G_{\phi^A}} &= \mathbb{E}_{\substack{\tilde{y}^A \sim p_{G_{\phi^A}}(y|x^B_0) \\ x_0^B \sim P_B}} \left[ -\log D_{\phi^A}(\tilde{y}^A) \right] \\
    L_{D_{\phi^A}} &= \mathbb{E}_{y^A \sim p_{A}}\left[ -\log D_{\phi^A}(y^A) \right] + \mathbb{E}_{\substack{\tilde{y}^A \sim p_{G_{\phi^A}}(y|x^B_0) \\ x_0^B \sim P_B}} \left[ -\log \left( 1 - D_{\phi^A}(\tilde{y}^A) \right) \right]
\end{split}
$}
\end{equation*}

\paragraph{Adversarial Loss for Diffusive Modules:} Uses non-saturating losses with a gradient penalty:
\begin{equation*}
\scalebox{0.8}{$
\begin{split}
    L_{G_{\theta^A}} &= \mathbb{E}_{
        \substack{
        \hat{x}^A_{t-1} \sim q\left(x^A_{t-1} | x^A_t,x_0 = G_{\theta^A}(x^A_t, y = \tilde{y}^B, t)\right),
         \\
        \tilde{y}^B \sim p_{G_{\phi^B}}(y|x^A_0), 
         \\
        x^A_{t} \sim q\left(x^A_{t} | x_0\right),
         \\
         t \sim \mathcal{U}\left( 1,T \right),
         \\
         x_0^A \sim P_A
        }
    }\left[ -\log D_{\theta^A}(\hat{x}^A_{t-1}) \right] \\
    L_{D_{\theta^A}} &= \mathbb{E}_{
            \substack{
            x^A_{t} \sim q\left(x^A_{t} | x_0\right),
             \\
             t \sim \mathcal{U}\left( 1,T \right),
             \\
             x_0^A \sim P_A
            }
        }
        \Bigg[\mathbb{E}_{\substack{
        x^A_{t-1} \sim q\left(x^A_{t-1} | x^A_t, x_0 = x^A_0\right)
        }
    } \left[-\log D_{\theta^A}(x^A_{t-1}) \right] \\
    &+ \mathbb{E}_{
        \substack{
        \hat{x}^A_{t-1} \sim q\left(x^A_{t-1} | x^A_t,x_0 = G_{\theta^A}(x^A_t, y = \tilde{y}^B, t)\right),
        \\
        \tilde{y}^B \sim p_{G_{\phi^B}}(y|x^A_0)
        \\
        }
    }\left[ -\log \left( 1 - D_{\theta^A}(\hat{x}^A_{t-1}) \right) \right]\\
    &+ \mathbb{E}_{\substack{
        x^A_{t-1} \sim q\left(x^A_{t-1} | x^A_t, x_0 = x^A_0\right)
        }
    } \left[\left\| D_{\theta^A}(x^A_{t-1}) \right\|_2^2 \right]
    \Bigg]
\end{split}
$}
\end{equation*}

\paragraph{Cycle-Consistency Loss:} Ensures the reconstruction of original images:
\begin{equation*}
\scalebox{0.8}{$
\begin{split}
    L_{\text{cycle}} = \mathbb{E}_{\substack{t \sim \mathcal{U}(1,T) \\ q(x_0^{A,B}) \\ q(x_t^{A,B}|x_0^{A,B}) \\ p_{G_{\phi^{A,B}}}(\tilde{y}^{B,A}|x^{A,B}_0) \\ p_{G_{\phi^{A,B}}}(\dot{x}^{A,B}_0 | \tilde{y}^{B,A}) \\ q(\tilde{x}_0^{A,B}|x_t^{A,B}, \tilde{y}^{B,A})}} \Bigg[ \lambda_{1 \phi} \left( \left| x_0^A - \dot{x}_0^A \right|_1 + \left| x_0^B - \dot{x}_0^B \right|_1 \right) + \lambda_{1 \theta} \left( \left| x_0^A - \tilde{x}_0^A \right|_1 + \left| x_0^B - \tilde{x}_0^B \right|_1 \right) \Bigg]
\end{split}
$}
\end{equation*}

\paragraph{Overall Objective:} Combines adversarial and cycle-consistency losses:
\begin{equation*}
\scalebox{0.8}{$
\begin{split}
    L_G &= L_{\text{cycle}} + \lambda_{2 \phi}\left( L_{G_{\phi^A}} + L_{G_{\phi^B}} \right) + \lambda_{2 \theta}\left( L_{G_{\theta^A}} + L_{G_{\theta^B}} \right) \\
    L_D &= \lambda_{2 \phi}\left( L_{D_{\phi^A}} + L_{D_{\phi^B}} \right) + \lambda_{2 \theta}\left( L_{D_{\theta^A}} + L_{D_{\theta^B}} \right)
\end{split}
$}
\end{equation*}

%% file: methods.tex
\section{Methods}

\begin{figure}[h]
\centering
\includegraphics[width=0.8\textwidth]{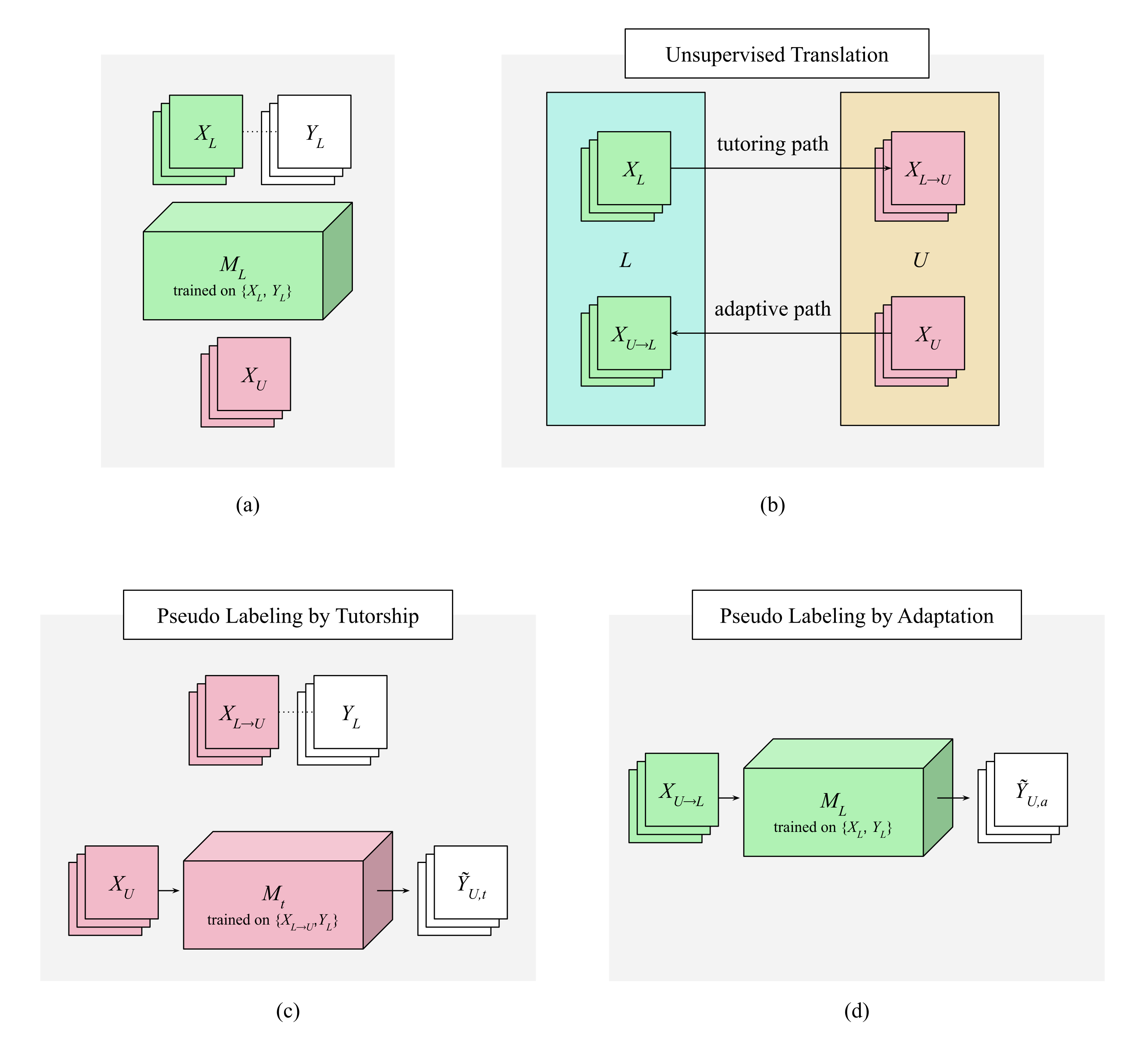}
% \includesvg[width=\textwidth]{figs/figure_methods.png}
\caption{Methodology overview. (a) Initial data and model (b) Unpaired translation between $L$ and $U$ via Adversarial Diffusion (c) Tutorship: Training a segmentation model on synthetic data $\{X_{L \rightarrow U}, Y_L\}$ for inference in $U$ (d) Adaptation: Applying a segmentation model pre-trained on $L$ to the translated images $X_{U \rightarrow L}$} \label{fig2-0}
\end{figure}
We have two similar image domains: $L$ (labeled) and $U$ (unlabeled). The task we perform on these two domains is shared --- in our case, segmentation of the axon and myelin tissues. Images $X_L$ sampled from $L$ are annotated with segmentation labels $Y_L$, whereas images $X_U$ sampled from $U$ are unlabeled. Typically, we have a model $M_L$ pre-trained on $\{X_L, Y_L\}$ but it doesn't generalize well to $U$. Our goal is ultimately to generate pseudo labels $\widetilde{Y}_U$ to bootstrap and accelerate the annotation of images $X_U$. 

As shown in Fig. \ref{fig2-0}b, our approach is to perform image translation between domains $L$ and $U$, resulting in synthetic images $X_{L \rightarrow U}$ and $X_{U \rightarrow L}$. We denote the translation directions $L \rightarrow U$ as \textit{tutoring path} and $U \rightarrow L$ as \textit{adaptive path}. Using the tutoring path, we obtain a synthetic dataset $\{ X_{L \rightarrow U}, Y_L \}$, effectively recycling the previously available annotations. A segmentation model is trained on this synthetic dataset and subsequently applied to $X_U$ to generate pseudo labels $\widetilde{Y}_{U,t}$. We denote this process pseudo labeling by \textit{tutorship}, because we use the annotations of data from $L$ to guide prediction on images from $U$, as illustrated in Fig. \ref{fig2-0}c.

$X_L$ and $X_U$ are unpaired, so we use an unsupervised image translation method based on cycle-consistency. Thus, although we are interested in the tutoring path yielding $X_{L \rightarrow U}$, we also get $X_{U \rightarrow L}$ from the adaptive path at no additional cost. We apply the pre-trained model $M_L$ on $X_{U \rightarrow L}$ to generate another set of pseudo labels $\widetilde{Y}_{U,a}$. We refer to this complementary process as pseudo labeling by \textit{adaptation}, because we are transferring the style of images $X_L$ (used to train $M_L$) onto images $X_U$, as shown in Fig. \ref{fig2-0}d.

\subsection{Data}
The 3 unlabeled datasets used in this project are described in Table \ref{data_table}, along with the reference annotated TEM dataset. Note the diversity in input modality (TEM/SEM/BF), species, pathology, body part, spatial resolution and median size. This diversity is illustrated in Fig. S1 in the Supplementary Material for a dataset preview. The \texttt{TEM} dataset represents the $X_L$ annotated images across all experiments. We evaluate our pipeline in 3 different settings, using in turn the \texttt{TEM-MACAQUE}, \texttt{SEM} and \texttt{BF} datasets to simulate the unannotated data pool $X_U$. In reality, these datasets are annotated so the pseudo labels can be evaluated quantitatively. Conveniently, the $X_L$ data pool has the smallest pixel size (highest magnification). As such, $X_L$ is resized to match the $X_U$ pixel size in all 3 experiments without significant loss of details.

\begin{table}
\centering
\caption{Dataset details}\label{data_table}
\begin{tabular}{l p{0.18\linewidth} p{0.18\linewidth} p{0.18\linewidth} p{0.18\linewidth}}
\hline
\textbf{dataset} & \texttt{TEM} & \texttt{TEM-MACAQUE} & \texttt{SEM} & \texttt{BF}\\
\hline
\hline
\textbf{species} & mouse & macaque & rat & rat, \\
 & & & & rabbit,\\
 & & & & human\\
\hline
\textbf{pathology} & healthy & healthy & healthy & healthy, \\
& & & &  demyelination \\
\hline
\textbf{organ} & brain & brain & spinal cord & brain,\\
& & & & spinal cord, \\
& & & & nerves, \\ 
& & & & muscles \\
\hline
\textbf{pixel size} & 0.00236 & 0.009 & 0.1 & 0.1 \\
(um/px)& & & & 0.211 \\
& & & & 0.226 \\
\hline
\textbf{median size} & 2286x3762 & 3040x2300 & 924x1278 & 1255x1440 \\
\hline
\textbf{nb. of images} & 158 & 16 & 10 & 39 \\
\hline
\end{tabular}
\end{table}

\subsection{Models}
\subsubsection{Image Translation} To perform unpaired modality translation, we utilized the SynDiff framework, which leverages conditional adversarial diffusion models for translation \cite{ozbey2023unsupervised}. Training on unpaired data is enabled by the use of a cycle consistency loss on a pair of non-diffusive GANs which generate initial unrefined estimates of the target translations. These GANs employ a 3-layer discriminator and a 6-block ResNet generator, both utilizing instance normalization, adapting the architecture from the pix2pix repository \cite{isola2017image}. Subsequently, these preliminary translations are passed as conditioning inputs to the denoising diffusion GANs, which are thus trained to produce translations between two modalities in an unpaired manner. At inference time, the non-diffusive modules are ignored, and the source image is directly used as the conditioning input for the denoising diffusion GANs.

The use of denoising diffusion GANs is motivated by resolving the trade-offs between the sampling efficiency of GANs and the training stability and mode coverage of diffusion models \cite{xiao2021tackling}. It has been demonstrated that with infinitesimally small time-steps, the reverse diffusion process mirrors the Gaussian nature of the forward diffusion process. Thus, if we model the forward diffusion process with a Gaussian distribution, we can assume the reverse process will also follow a Gaussian distribution \cite{ho2020denoising, sohl2015deep, xiao2021tackling}. As such, one major limitation of standard diffusion models is their requirement for a large number of timesteps with minimal variance steps for this assumption to hold, significantly increasing sampling time. This is suboptimal for tasks such as translating whole slide images. In contrast, GANs generate high-quality samples with low sampling times but often suffer from reduced sample diversity due to issues like mode collapse \cite{salimans2016improved, zhao2018bias}. 

Denoising diffusion GANs combine the advantages of both approaches—high quality and sampling speed with extensive mode coverage—by bypassing the assumption of normally distributed data at each diffusion step and using fewer than 8 timesteps. To acheive this, they employ GANs to model transitions between each denoising step \cite{xiao2021tackling}. It can be shown, based on the formulation by Ho et al., that the original DDPM can be equivalent to first predicting \(x_0\) and then reapplying noise via \(p_{\theta}\left( x_{t-1}|x_t \right) = q(x_{t-1}|x_t, x_0=f_{\theta}(x, t))\). Denoising diffusion GANs parameterize \(f_{\theta}\) using GANs instead of deterministically predicting the original image \(x_0\) \cite{xiao2021tackling}. Specifically, SynDiff parameterizes \(f_{\theta}\) using the NCSN++ generator \cite{song2020score}, with a U-net structure featuring ResNet and Attention blocks, z-conditioning via mapping networks, and adaptive group normalization, while its time-dependent discriminator employs a convolutional ResNet architecture, sinusoidal positional embeddings, and minibatch standard deviation layers, following the implementation described by Xiao et al. \cite{xiao2021tackling}. For more details on the training objective, please refer to the supplementary materials. 

To evaluate SynDiff checkpoints, we used SSIM \cite{wang2004image, hore2010image}, PSNR \cite{hore2010image}, and \(L1\)-loss. These metrics, requiring ground truths, were computed after a full \(L \rightarrow U \rightarrow L\) translation cycle, comparing the reconstruction \(X_{L \rightarrow U \rightarrow L}\) with the original labeled image \(X_L\).

\subsubsection{Image Segmentation} 
The nnU-Net framework has been selected for our study due to its demonstrated robustness and superior performance in medical image segmentation tasks. Its automated configuration and adaptability make it particularly effective for small and diverse medical datasets. nnU-Net’s design principles ensure rigorous validation and optimal architectural choices, consistently delivering state-of-the-art results \cite{isensee2018nnu, isensee2024nnu}. nnU-Net employs a 5-fold cross-validation approach, training five different models on five different folds. The best checkpoint from each fold is then used to compute the standard deviation of the segmentation scores.

\subsection{Experiments}
\begin{figure}
\centering
\includegraphics[width=0.8\textwidth]{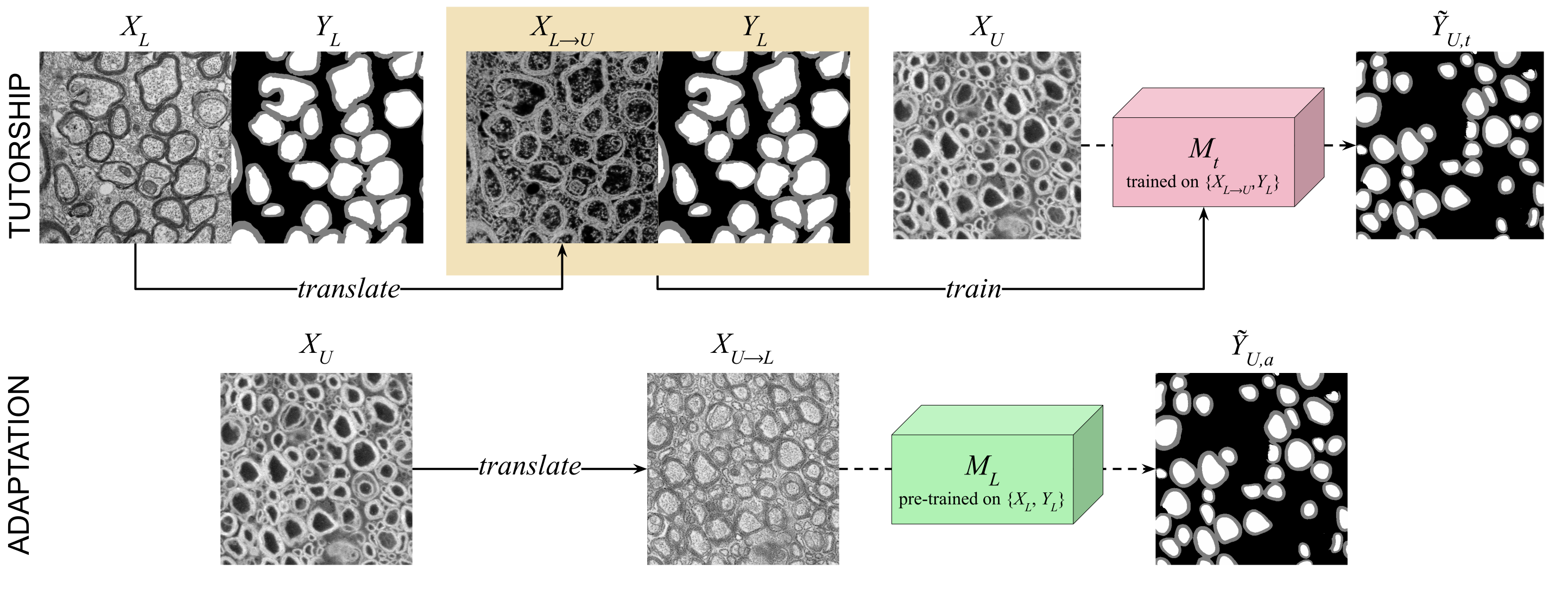}
% \includesvg[width=\textwidth]{figs/pseudo_labeling_example_sem_v2.svg}
\caption{Overview of the two pseudo labeling strategies using TEM as the labeled dataset and SEM as the unlabeled data. (Top) The Tutorship strategy involves generating a synthetic dataset by translating labeled TEM images and training a proxy segmentation model on this synthetic data. (Bottom) The Adaptation strategy converts unlabeled SEM images to match the TEM data distribution, enabling segmentation with a pre-trained model.} \label{fig2-sem-example}
\end{figure}

The following 3 experiments are increasing challenging, as the domains $L$ and $U$ are progressively more dissimilar. In each case, we evaluate pseudo labeling strategies using the tutoring and adaptive paths. Both methods are illustrated in Fig. \ref{fig2-sem-example}.

\subsubsection{TEM $\leftrightarrow$ TEM-MACAQUE} In this experiment, both $X_L$ and $X_U$ were acquired using transmission electron microscopy at similar magnifications. The main difference between these datasets is the provenance of the samples in terms of species, and a slightly different intensity profile (most likely due to sample preparation and acquisition parameters). This experiment acts as a baseline and shows how our pseudo labeling method performs on a small domain shift.

\subsubsection{TEM $\leftrightarrow$ SEM}
Here, $X_L$ and $X_U$ were acquired with different modalities. \texttt{TEM} and \texttt{SEM} have opposite contrasts (i.e. myelin is dark in TEM, white in SEM. See Supplemental Materials, Fig. S1). The species, body part and pixel size are also different, resulting in an intermediate domain shift.

\subsubsection{TEM $\leftrightarrow$ BF}
This final experiment investigates a larger domain shift between $X_L$ and $X_U$. The \texttt{BF} dataset exhibits a wide diversity across 4 organs, including 3 different species, pathology and pixel size.

%% file: results.tex
\section{Results and Discussion}
\begin{figure}
\centering
\includegraphics[width=0.8\textwidth]{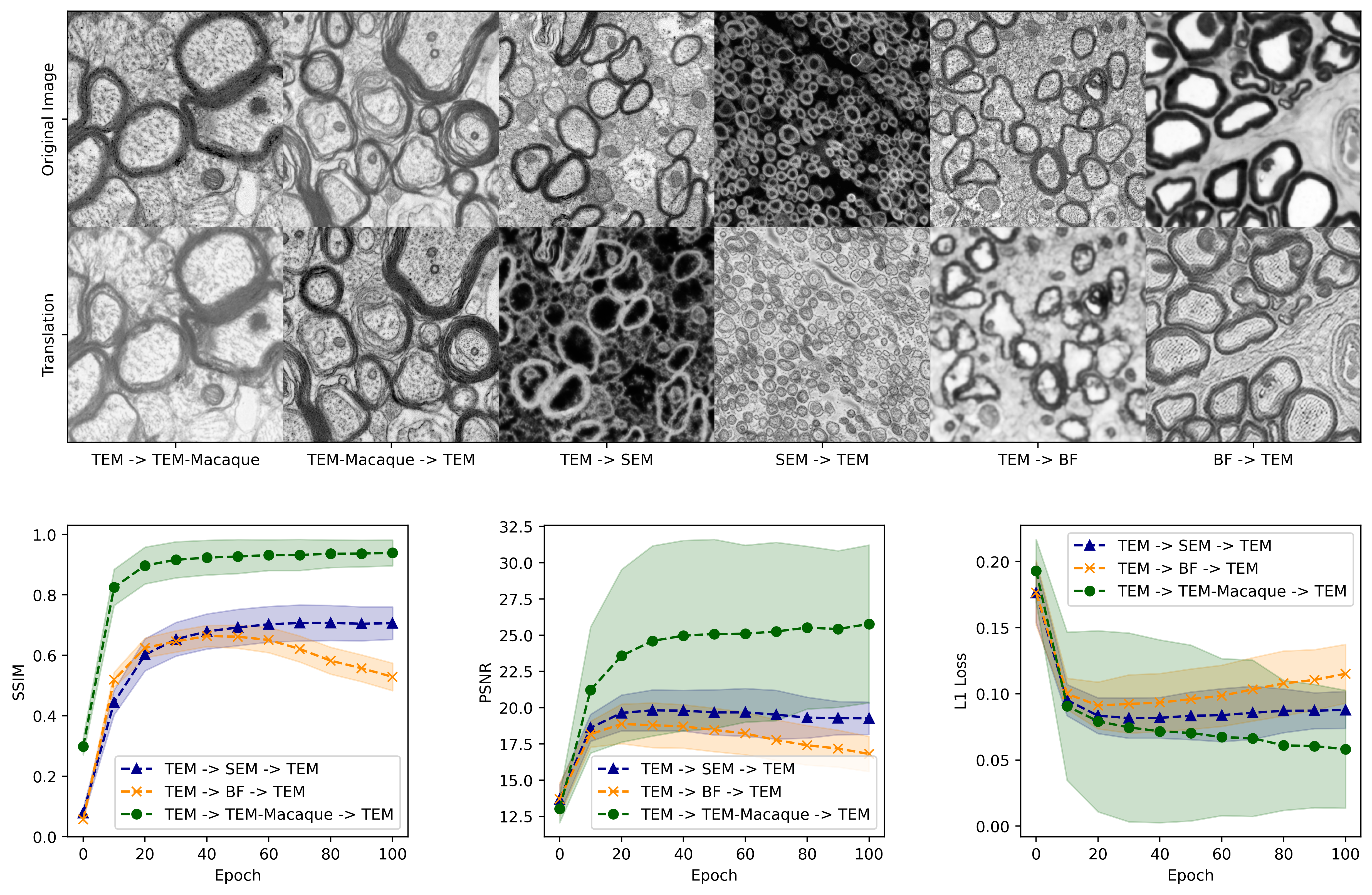}
\caption{(Top) Image translation examples. (Bottom) Unsupervised translation reconstruction losses: Structural Similarity Index Measure (SSIM), Peak Signal-to-Noise Ratio (PSNR) and \(L1\). The error bars represent the standard deviations among the validation images.} \label{fig3-losses}
\end{figure}

Learning curves featuring the peak signal-to-noise ratio (PSNR), the structural similarity index measure (SSIM), and the \(L1\)-loss of the image translation models after an \(L \rightarrow U \rightarrow L\) translation cycle are reported in Fig. \ref{fig3-losses}. These curves, along with visual assessment of the generation process, guided the selection of the best checkpoints. Specifically, checkpoints for TEM $\leftrightarrow$ TEM-MACAQUE, TEM $\leftrightarrow$ SEM, and TEM $\leftrightarrow$ BF were selected at epochs 100, 50, and 40, respectively.

\begin{figure}
\centering
\includegraphics[width=0.8\textwidth]{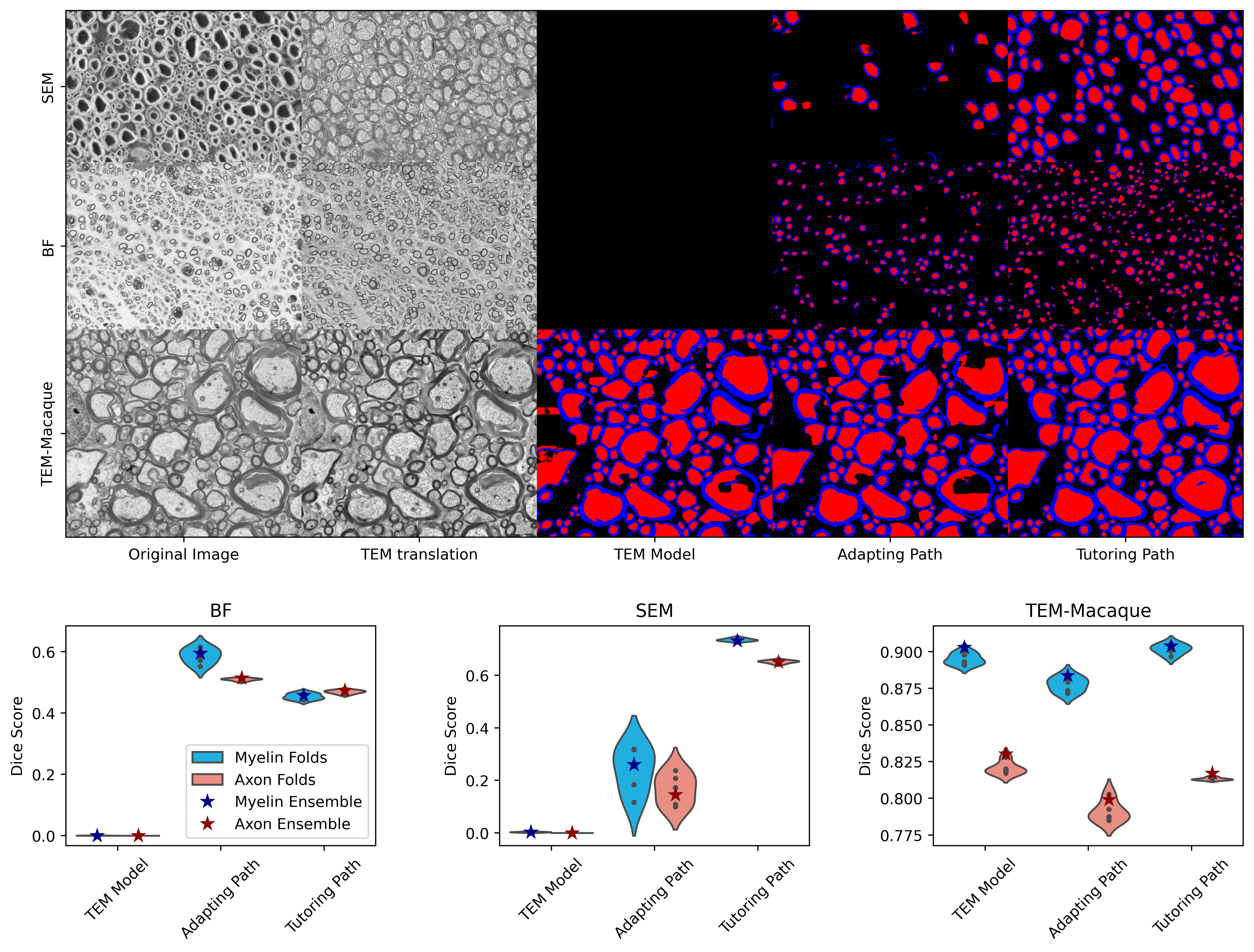}
\caption{Pseudo labeling results. (Top) Original and translated image alongside the pre-trained model prediction and pseudo labels. (Bottom) Dice scores under 5-fold cross-validation.} \label{fig3-pseudolabels}
\end{figure}

As illustrated in Fig. \ref{fig3-pseudolabels}, our pseudo labeling strategy is effective, especially using the tutoring path. The Dice score, sensitivity, and specificity computed on the pseudo labels, along with the standard deviations of each metric across all five folds of the nnU-Net segmentations, are reported in Table \ref{table3}. For the TEM $\leftrightarrow$ TEM-MACAQUE experiment, using the pre-trained model provides a similar performance to our pseudo labeling strategies. This highlights the fact that when $X_L$ and $X_U$ are sufficiently similar, using a model pre-trained on $X_L$ is more than enough to produce good pseudo labels. However, this is not the case for the two remaining experiments, where the pre-trained TEM model performance drops to zero, as expected due to larger domain shifts. Both the TEM $\leftrightarrow$ SEM and TEM $\leftrightarrow$ BF experiments demonstrate the benefit of our pseudo labeling strategies. In the TEM $\leftrightarrow$ SEM case, the tutoring path provides a better prediction than the adapting path, the latter having a much lower sensitivity. Using the tutorship method, the mean Dice score averages \(0.736 \pm 0.005\) for axons and \(0.652 \pm 0.005\) for myelin, a remarkable starting point in the context of manual annotation. In the TEM $\leftrightarrow$ BF scenario, the adapting path marginally surpasses the tutoring path, achieving a mean Dice score of \(0.586 \pm 0.025\) for axons and \(0.510 \pm 0.004\) for myelin. While further research is required to elucidate the conditions under which the tutoring path outperforms the adapting path and vice versa, it is plausible that the performance is contingent on the efficacy of the translation model in either the adapting \(U \rightarrow L\) or tutoring \(L \rightarrow U\) direction.

\begin{table}[]
\caption{Pseudo Label Dice Score, Specificity and Sensitivity. TEM-M stands for the TEM-MACAQUE dataset. Applying the pre-trained TEM model directly on the SEM/BF datasets produced empty masks, so we omit these cases.}
\label{table3}
\centering
\begin{tabular}{cclcll}
\multicolumn{1}{c}{Class} &
  $X_U$ &
  \multicolumn{1}{c}{Method} &
  Dice Score &
  \multicolumn{1}{c}{Specificity} &
  \multicolumn{1}{c}{Sensitivity} \\ \hline
\multicolumn{1}{|c|}{\multirow{7}{*}{Axon}} &
  \multicolumn{1}{c|}{\multirow{3}{*}{TEM-M}} &
  \multicolumn{1}{l|}{TEM model} &
  \multicolumn{1}{c|}{0.895 ± 0.004} &
  \multicolumn{1}{l|}{0.935 ± 0.007} &
  \multicolumn{1}{l|}{0.905 ± 0.015} \\ 
\multicolumn{1}{|c|}{} &
  \multicolumn{1}{c|}{} &
  \multicolumn{1}{l|}{Adaptation} &
  \multicolumn{1}{c|}{0.878 ± 0.005} &
  \multicolumn{1}{l|}{0.936 ± 0.006} &
  \multicolumn{1}{l|}{0.872 ± 0.017} \\  
\multicolumn{1}{|c|}{} &
  \multicolumn{1}{c|}{} &
  \multicolumn{1}{l|}{Tutorship} &
  \multicolumn{1}{c|}{\textbf{0.902 ± 0.003}} &
  \multicolumn{1}{l|}{0.932 ± 0.001} &
  \multicolumn{1}{l|}{0.922 ± 0.006} \\ \cline{2-6} 
\multicolumn{1}{|c|}{} &
  \multicolumn{1}{c|}{\multirow{2}{*}{SEM}} &
  \multicolumn{1}{l|}{Adaptation} &
  \multicolumn{1}{c|}{0.238 ± 0.088} &
  \multicolumn{1}{l|}{0.984 ± 0.010} &
  \multicolumn{1}{l|}{0.144 ± 0.061} \\ 
\multicolumn{1}{|c|}{} &
  \multicolumn{1}{c|}{} &
  \multicolumn{1}{l|}{Tutorship} &
  \multicolumn{1}{c|}{\textbf{0.736 ± 0.005}} &
  \multicolumn{1}{l|}{0.914 ± 0.002} &
  \multicolumn{1}{l|}{0.705 ± 0.009} \\ \cline{2-6} 
\multicolumn{1}{|c|}{} &
  \multicolumn{1}{c|}{\multirow{2}{*}{BF}} &
  \multicolumn{1}{l|}{Adaptation} &
  \multicolumn{1}{c|}{\textbf{0.586 ± 0.025}} &
  \multicolumn{1}{l|}{0.949 ± 0.008} &
  \multicolumn{1}{l|}{0.889 ± 0.022} \\ 
\multicolumn{1}{|c|}{} &
  \multicolumn{1}{c|}{} &
  \multicolumn{1}{l|}{Tutorship} &
  \multicolumn{1}{c|}{0.454 ± 0.010} &
  \multicolumn{1}{l|}{0.915 ± 0.003} &
  \multicolumn{1}{l|}{0.942 ± 0.003} \\ \hline
\multicolumn{1}{|c|}{\multirow{7}{*}{Myelin}} &
  \multicolumn{1}{c|}{\multirow{3}{*}{TEM-M}} &
  \multicolumn{1}{l|}{TEM model} &
  \multicolumn{1}{c|}{\textbf{0.820 ± 0.004}} &
  \multicolumn{1}{l|}{0.945 ± 0.009} &
  \multicolumn{1}{l|}{0.811 ± 0.021} \\ 
\multicolumn{1}{|c|}{} &
  \multicolumn{1}{c|}{} &
  \multicolumn{1}{l|}{Adaptation} &
  \multicolumn{1}{c|}{0.791 ± 0.007} &
  \multicolumn{1}{l|}{0.938 ± 0.005} &
  \multicolumn{1}{l|}{0.777 ± 0.022} \\ 
\multicolumn{1}{|c|}{} &
  \multicolumn{1}{c|}{} &
  \multicolumn{1}{l|}{Tutorship} &
  \multicolumn{1}{c|}{0.813 ± 0.001} &
  \multicolumn{1}{l|}{0.934 ± 0.000} &
  \multicolumn{1}{l|}{0.822 ± 0.001} \\ \cline{2-6} 
\multicolumn{1}{|c|}{} &
  \multicolumn{1}{c|}{\multirow{2}{*}{SEM}} &
  \multicolumn{1}{l|}{Adaptation} &
  \multicolumn{1}{c|}{0.165 ± 0.060} &
  \multicolumn{1}{l|}{0.995 ± 0.003} &
  \multicolumn{1}{l|}{0.092 ± 0.036} \\ 
\multicolumn{1}{|c|}{} &
  \multicolumn{1}{c|}{} &
  \multicolumn{1}{l|}{Tutorship} &
  \multicolumn{1}{c|}{\textbf{0.652 ± 0.005}} &
  \multicolumn{1}{l|}{0.962 ± 0.002} &
  \multicolumn{1}{l|}{0.512 ± 0.008} \\ \cline{2-6} 
\multicolumn{1}{|c|}{} &
  \multicolumn{1}{c|}{\multirow{2}{*}{BF}} &
  \multicolumn{1}{l|}{Adaptation} &
  \multicolumn{1}{c|}{\textbf{0.510 ± 0.004}} &
  \multicolumn{1}{l|}{0.949 ± 0.003} &
  \multicolumn{1}{l|}{0.635 ± 0.020} \\ 
\multicolumn{1}{|c|}{} &
  \multicolumn{1}{c|}{} &
  \multicolumn{1}{l|}{Tutorship} &
  \multicolumn{1}{c|}{0.469 ± 0.005} &
  \multicolumn{1}{l|}{0.947 ± 0.002} &
  \multicolumn{1}{l|}{0.566 ± 0.012} \\ \hline
\end{tabular}
\end{table}